\documentclass[sigconf, screen]{acmart}
\usepackage{enumitem}
\usepackage{bm}
\usepackage{multirow} 
\usepackage{amsmath}
\usepackage{etoolbox}
\usepackage{adjustbox}
\usepackage{tcolorbox}
\usepackage{colortbl}
\usepackage{geometry}
\usepackage{balance}
\newcommand{\eg}{\emph{e.g., }}

\AtBeginDocument{%
  }

\setcopyright{acmlicensed}
\copyrightyear{2018}
\acmYear{2018}
\acmDOI{XXXXXXX.XXXXXXX}
\acmConference[Conference acronym 'XX]{Make sure to enter the correct
  conference title from your rights confirmation email}{June 03--05,
  2018}{Woodstock, NY}
\acmISBN{978-1-4503-XXXX-X/2018/06}

\begin{document}
\begin{sloppypar}

\title{Large Language Model Empowered Recommendation Meets All-domain Continual Pre-Training}

\author{Haokai Ma}
\email{haokai.ma1997@gmail.com}
\affiliation{
  \institution{National University of Singapore}
  \country{Singapore}
}

\author{Yunshan Ma}
\email{ysma@smu.edu.sg}
\affiliation{
  \institution{Singapore Management University}
  \country{Singapore}
}

\author{Ruobing Xie}
\authornote{Corresponding author.}
\email{xrbsnowing@163.com}
\affiliation{
  \institution{Tencent}
  \country{China}
}

\author{Lei Meng}
\email{lmeng@sdu.edu.cn}
\affiliation{
  \institution{Shandong University}
  \country{China}
}

\author{Jialie Shen}
\email{jerry.shen@city.ac.uk}
\affiliation{
  \institution{City, University London}
  \country{United Kingdom}
}

\author{Xingwu Sun}
\email{sunxingwu01@gmail.com}
\affiliation{
  \institution{Tencent}
  \country{China}
}

\author{Zhanhui Kang}
\email{kegokang@tencent.com}
\affiliation{
  \institution{Tencent}
  \country{China}
}

\author{Tat-Seng Chua}
\email{dcscts@nus.edu.sg}
\affiliation{
  \institution{National University of Singapore}
  \country{Singapore}
}

\renewcommand{\shortauthors}{Haokai Ma et al.}

\begin{abstract}
Recent research efforts have investigated how to integrate Large Language Models (LLMs) into recommendation, capitalizing on their semantic comprehension and open-world knowledge for user behavior understanding. These approaches predominantly employ supervised fine-tuning on single-domain user interactions to adapt LLMs for specific recommendation tasks. However, they typically encounter dual challenges: the mismatch between general language representations and domain-specific preference patterns, as well as the limited adaptability to multi-domain recommendation scenarios. To bridge these gaps, we introduce CPRec—an all-domain Continual Pre-training framework for Recommendation—designed to holistically align LLMs with universal user behaviors through the continual pre-training paradigm. Specifically, we first design a unified prompt template and organize users’ multi-domain behaviors into domain-specific behavioral sequences and all-domain mixed behavioral sequences that emulate real-world user decision logic. To optimize behavioral knowledge infusion, we devise a Warmup-Stable-Annealing learning rate schedule tailored for the continual pre-training paradigm in recommendation to progressively enhance the LLM's capability in knowledge adaptation from open-world knowledge to universal recommendation tasks. To evaluate the effectiveness of our CPRec, we implement it on a large-scale dataset covering seven domains and conduct extensive experiments on five real-world datasets from two distinct platforms. Experimental results confirm that our continual pre-training paradigm significantly mitigates the semantic-behavioral discrepancy and achieves state-of-the-art performance in all recommendation scenarios. The source code will be released upon acceptance.
\end{abstract}



\maketitle
\section{Introduction} \label{sec:introduction}
Recommender system (RS) aims to predict the appropriate next-item according to users' historical behaviors~\cite{LLM_SR_1,CL4SRec,MCDRec}. However, current RS approaches typically encounter the data sparsity problem since users usually have few behaviors~\cite{LLM_SR_2, PDRec,CIERec}. Zooming into this issue, the multi-modal knowledge and the cross-domain transition become the promising solutions by leveraging the additional information from other modalities and domains, which have gained increasing research attention recently\cite{Recformer, Tri-CDR}.

The emergence of Large Language Models (LLMs) brings new opportunities in solving the data sparsity issue in RS~\cite{LLM_RS_1,LLM_RS_2}. Existing studies typically explore LLM for recommendation (LLM4Rec) from the following three aspects: LLM-as-encoder, LLM-as-reasoner, and LLM-as-recommender. 
Specifically, LLM-as-encoder aims to encode user profiles and item descriptions into the latent space via LLMs to improve recommenders \cite{KAR}. While capable of modeling complex contextual information, it struggles to mine domain-specific knowledge and is constrained by the capability limits of its backbone.
As for LLM-as-reasoner, it directly inputs historical interactions into LLMs to infer user preferences~\cite{ChatRec,LLMRank}. However, the open-world knowledge in LLMs struggles to effectively generalize to personalized user preferences, particularly in capturing dynamic preference shifts.
LLM-as-recommender prefers adapting LLMs to solve preference understanding tasks with historical behaviors under the supervised fine-tuning (SFT) paradigm~\cite{InstructRec, TALLRec, LLaRA}. Moreover, some works further employ Direct Preference Optimization (DPO) to explicitly optimize LLMs for personalized ranking via users' pair-wise preferences~\cite{S-DPO,RosePO,LLM_RS_3}. 
Nevertheless, \emph{\textbf{concerning the intrinsic discrepancy between semantic understanding and collaborative comprehension, aligning LLMs with limited domain-specific behaviors in a one-step manner is not perfectly satisfactory.}}

\begin{figure}[!t]
    \centering
    \includegraphics[width=0.98\linewidth]{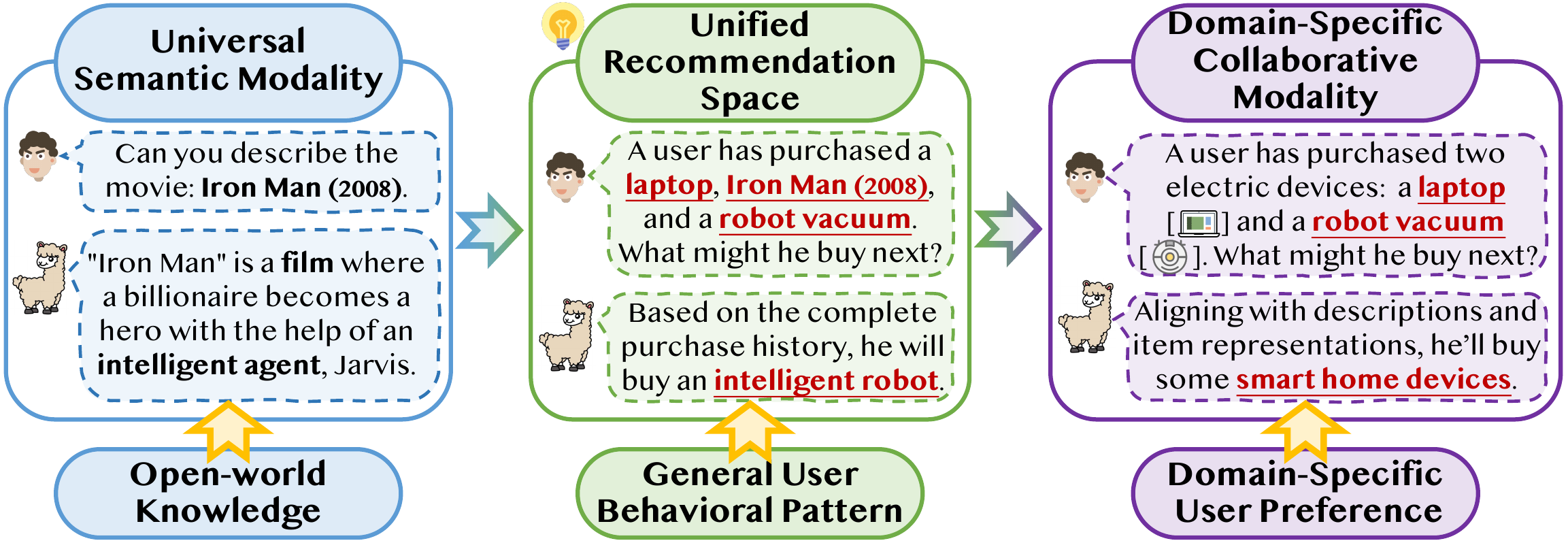}
    \vspace{-0.2cm}
    \caption{The modality adaptation pathway in our work. It progressively shifts the open-world knowledge from the universal semantic modality to the specific collaborative modality via the general user behavioral patterns embedded in our all-domain continual pre-training paradigm.}
    \vspace{-0.4cm}
    \label{fig.motivation}
\end{figure}

Although these studies demonstrate the potential of LLMs in RS, there are still multiple research gaps in alleviating the gap between semantic and collaborative modalities and leveraging data from various other domains:
(1) LLMs are typically pre-trained on the large-scale textual data collected from Internet to model the open-world semantic knowledge and fundamental competencies, while RS require personalized collaborative information, which is not inherently preserved by LLMs. This implies the \textbf{\emph{significant modality gap}} between open-world semantic knowledge and general preference patterns and the necessity of a meticulous modality adaptation from the universal semantic modality to the specific collaborative modality. 
(2) Some previous endeavors \cite{Recformer, UniSRec} have explored the multi-domain pre-training for RS through the \textbf{\emph{fragmented behavior modeling}} on user behaviors from multiple domains, overlooking the mixed behavioral sequences that serve as the authentic user behavioral logic. Moreover, they are commonly built on the relatively smaller language models, limiting their ability to harness the potential information gains from LLMs.
(3) The advanced LLM4Rec studies \cite{LLaRA,CoRA} predominantly conduct SFT with the single-domain user interactions, making them implausibly to scale effectively to multi-domain settings. This \textbf{\emph{insufficient multi-domain generalization}} further impacts the comprehensive user preference modeling and leads to suboptimal recommendation performance. This issue is particularly pronounced in few-shot RS scenarios. 

To mitigate these limitations, this work attempts to establish a general framework targeted at all-domain \textbf{C}ontinual \textbf{P}re-training for \textbf{Rec}ommendation (CPRec). Specifically, we aim to introduce the continual pre-training (CPT) paradigm into the existing LLM4Rec methods, thereby integrating the users' behavioral patterns from multiple available domains into LLMs before SFT. As illustrated in Figure~\ref{fig.motivation}, \textbf{\emph{this paradigm shift facilitates the smooth adaptation from the open-world semantic modality into the domain-specific collaborative modality through a bridging unified recommendation space.}}
To solve the fragmented behavior modeling problem, we first gather user behaviors from multiple available domains and chronologically align them, yielding a comprehensive data that well perceives users' authentic behavioral patterns.
Subsequently, we re-formulate these behavioral sequences with a domain-agnostic prompt template to enable LLMs to acquire the more generalized behavior modeling capability, thereby alleviating the significant distribution gap issue. This ensures the smooth modality adaptation during the CPT while remaining consistent with the original optimization objective of LLMs.
Finally, to continuously fine-tune the pre-trained LLMs to better fit the recommendation task, we propose a Warmup-Stale-Annealing (WSA) learning rate scheduler. That is, we incorporate the domain-specific behaviors and all-domain mixed behaviors (if exist) into CPT to address the insufficient multi-domain generalization issue. Following these technical details, we construct the generalized LLMs in the unified recommendation space and then initialize the existing LLM-enhanced recommenders to achieve improvements in both recommendation performance and robustness of multiple domains.

We conduct extensive experiments on five real-world datasets from two platforms to evaluate our CPRec. The experimental results, including performance comparison, ablation studies, universality analysis, and in-depth analysis, verify its effectiveness and robustness. The contributions are summarized as follows:
\begin{itemize}[leftmargin=*, topsep=0.2pt,parsep=0pt]
  \item We have verified the necessity of all-domain continual pre-training in smoothly adapting open-world semantic knowledge within LLMs into specific collaborative modality. To the best of our knowledge, we are among the first to explore the all-domain continual pre-training paradigm in LLM4Rec.
  \item We examine the optimal configurations of data composition, prompt template, and training scheduler in CPRec and provide several practical informative insights about this paradigm.
  \item Extensive experiments demonstrate the effectiveness, robustness, and universality of CPRec on various recommendation scenarios, even on zero-shot scenarios without SFT on target domains. CPRec could be the fundamental step of LLM4Rec.
\end{itemize}

\section{Related Work}\label{sec:related_work}
\subsection{LLMs for Recommendation}
LLMs have exhibited remarkable capability in context understanding \cite{LLM_understanding_1} and commonsense reasoning \cite{LLM_reasoning_1, LLM_reasoning_2}, revolutionizing the learning paradigm of recommendation. Different from the mainstreaming ID paradigm in traditional recommenders, LLM4Rec attempt to explore LLMs from three aspects: LLM-as-encoder, LLM-as-reasoner, and LLM-as-recommender.
For LLM-as-encoder, KAR \cite{KAR} jointly integrates the reasoned preferences knowledge and factual knowledge into RS via hybrid-expert adaptor. 
In contrast, LLM-as-reasoner studies (\eg ChatRec \cite{ChatRec} and LLMRank \cite{LLMRank}) design the prompt templates to incorporate user behaviors into closed-source LLMs and obtain ideal zero-shot performance. 
Regarding LLM-as-recommender, InstructRec \cite{InstructRec} and TALLRec \cite{TALLRec} convert users' historical behaviors into plaintext to align LLMs with the specific recommendation tasks. LLaRA \cite{LLaRA} is one of the SOTA LLM4Rec works, which proposes a hybrid prompt strategy to incorporate pre-extracted collaborative signals with the textual metadata and optimize LLMs to familiarize the RS mechanism with a curriculum tuning function.

However, these LLM-enhanced recommenders are commonly tailored for the domain-specific recommendation task, which exhibits poor flexibility and suboptimal performance in other recommendation scenarios. Meanwhile, concerning the intrinsic discrepancy between semantic understanding and collaborative awareness tasks, aligning the pre-trained LLMs with behavioral data within the recommendation task in a one-step manner is non-trivial. Therefore, an intermediate bridge, which is represented by the unified user preference patterns, is necessary for facilitating the semantic-to-collaborative smooth adaptation.

\subsection{Continual Pre-training}
Data-driven LLMs typically exhibit significant performance degradation in other domains when trained on a new downstream task, referred to as "Catastrophic Forgetting" \cite{CT-LLM}. This prompts the investigation of CPT, which adapts pre-trained LLMs to other unseen domains via the additional training on the newly-collected data \cite{CPT_3}. A series of studies is dedicated to mitigating the distribution shift issue caused by the post-training of new data while conserving computational resources \cite{CPT_1,CPT_2}. However, these researches are impossible to be directly applied to the recommendation tasks, as the complex collaborative pattern within user sequential behaviors exhibits significant deviations in natural language formats. To be specific, the differences between our CPRec with existing CPT approaches are three-fold: 
(1) In contrast to overly emphasizing data allocation for complex reasoning tasks, we leverage user behaviors from multiple domains in CPT to collaborate with the recommendation characteristics.
(2) We innovatively incorporate user behaviors with diverse hardness in preference understanding into distinct training stages with a tailored learning rate scheduler, smoothly adapting the semantic knowledge within pre-trained LLMs to collaborative modality.
(3) CPRec is a domain-agnostic approach that can be flexibly applied to diverse recommendation datasets from various domains and platforms, achieving consistent performance improvements.

\section{Preliminary}\label{sec:preliminary}
We present the base LLM-enhanced recommender in our CPRec and the empirical study about the drawbacks within the specific cross-domain transition in existing LLM4Rec studies.
\subsection{Base LLM-enhanced Recommender} 
We apply LLaRA \cite{LLaRA} as our base LLM-enhanced recommender for the diverse downstream recommendation tasks. It first represents item $i$ with the combination of its textual token $\bm{e}^t_i$ from the pre-trained LLM (\eg Llama2-7B \cite{Llama2}) and the pre-extracted collaborative representation $\bm{e}^c_i$ from the traditional recommender (\eg SASRec \cite{SASRec}), which can be formulated as: 
\vspace{-0.1cm}
\begin{equation}
\bm{e}_i\!=\!\operatorname{Concat}(\bm{e}^t_i, \bm{e}^c_i)\!=\!\operatorname{Concat}(\operatorname{LLM}_{\mu+\gamma}(t_i), \operatorname{Proj}(\operatorname{SAS}_{\theta}(i))),
\vspace{-0.05cm}
\end{equation}
\noindent
where $\bm{e}_i$ and $t_i$ denote the final representation and the semantic metadata of item $i$, $\operatorname{Proj}(.)$ is the trainable projector to project the collaborative features into the semantic space, $\mu$ denotes the frozen parameters of pre-trained LLM, and $\gamma$ denotes the trainable parameters under the Low-Rank Adaptation (LoRA) approach, $\operatorname{LLM}_{\mu+\gamma}(.)$ and $\operatorname{SAS}_{\theta}(.)$ denote the tokenizer of Llama \cite{Llama2} and the embedding layer of SASRec \cite{SASRec} respectively. With this combination, $\bm{e}_i$ can comprehensively represent item $i$ within both the semantic and collaborative aspects.

Specifically, we leverage a hybrid prompt to incorporate the collaborative representation $\bm{e}^c_i$ into the text-only prompt, bringing the integration of user behaviors and the textual understanding capability of LLMs. Finally, we optimize the LLM from the text-only prompts (easier) to the hybrid prompts (harder) in a curriculum learning paradigm. Here, we regards $p\!=\!\frac{\tau}{T}, (0\!\leq\!\tau\!\leq\!T)$ as the probability of modeling the hybrid prompts at step $\tau$, where $T$ denotes the whole steps in one epoch. 
It implies that the difficulty of the prompts used for tuning steadily grows until they stabilize as the hybrid prompts, which aligns perfectly with its inference process. This curriculum tuning strategy can facilitate LLMs gradually familiarizing themselves with recommendation tasks while avoiding being overwhelmed by complex tasks.

\subsection{Empirical Analysis of Drawbacks within Specific Cross-domain Transition}
The extensive open-world knowledge and advanced textual reasoning ability inherent in LLMs have gained broad recognition from both academia and industry. Theoretically, a well-trained LLM can handle various downstream tasks. Nevertheless, whether such semantic understanding ability can be seamlessly adapted to collaborative comprehension task is yet to be thoroughly validated. To investigate it, we conduct an intuitive experiment to compare the performance between SASRec, LLaRA w/o SFT, CD-LLaRA w/o SFT, CD-LLaRA, and LLaRA. Here, LLaRA w/o SFT only leverages the pre-trained LLaMA-7B and item representations to infer user preference, and CD-LLaRA denotes leveraging the LLM checkpoint pre-trained on another Music dataset to initialize the low-rank decomposition matrices in LLaRA.

\begin{figure}[!t]
    \centering
    \includegraphics[width=0.98\linewidth]{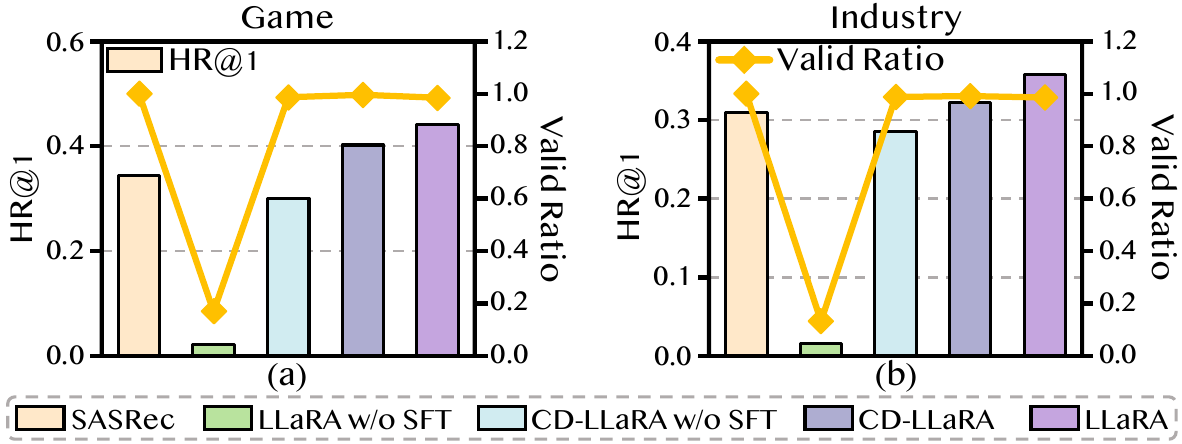}
    \vspace{-0.3cm}
    \caption{Results of SASRec, LLaRA w/o SFT, CD-LLaRA w/o SFT, CD-LLaRA, and LLaRA on two datasets. Removing the tailored SFT process results in a cliff-like performance drop, implying the notable gap between open-world semantic knowledge in LLM and collaborative information in RS. A simple cross-domain pre-trained LLM can alleviate it while there is still room for further improvement.}
    \vspace{-0.3cm}
    \label{fig.empirical_analysis}
\end{figure}

From Figure~\ref{fig.empirical_analysis}, we can observe that: 
(1) LLaRA outperforms all the compared methods, indicating that LLMs' remarkable reasoning capability can be effectively leveraged to support collaborative understanding tasks with the tailored SFT process.
(2) LLaRA w/o SFT even exhibits performance inferior to that of random selection, highlighting a typical drawback of LLM in recommendation: \textbf{\emph{the foundational distributions within LLMs deviate notably from user preference patterns}}.
(3) Benefiting from the collaborative-aware initialization of a pre-trained LLM on other domain, CD-LLaRA w/o SFT achieves comparable performance to SASRec without any time-consuming SFT process. \textbf{\emph{It confirms the effectiveness of an intermediate space in bridging general semantic modality and specific collaborative modality to some extent}}.
(4) CD-LLaRA exhibits inferior performance than LLaRA while exceeding all other methods, which may due to the fact that the pre-trained LLM collapses into a singular position that overly aligns with user preferences specific to the Music dataset. Inspired by the pre-training process of LLMs, we argue that \textbf{\emph{(more extensive and diverse) all-domain pre-training on user behaviors is essential to improve collaborative understanding ability of LLMs}}.
To sum up, these discoveries prompt us to explore the all-domain continual pre-training paradigm tailored for RS, enabling the smooth adaptation from LLMs' open-world semantic modality to the specific collaborative modality, thereby enhancing their utility in LLM4Rec.
\section{Methodology}\label{sec:methodology}
\subsection{Task Formulation}
CPRec aims to design a universal framework to smoothly adapt the open-world semantic knowledge within LLMs into specific collaborative modality via the all-domain continual pre-training paradigm. To achieve this, we construct user behavioral sequences and the corresponding textual metadata matrices for multiple domains. Without losing generality, we take domain $A$ as an example. We first define the behavior sequence of user $u$ in domain $A$ as $S^{A}_u\!=\!\{i^A_1, i^A_2, \cdots, i^A_p\}$, where $i^A_j \!\in\! \mathcal{I}^A$ is the $j$-th interacted item of user $u$ in domain $A$, and $p$ is the number of historical behaviors. Besides, we also pre-extract the corresponding textual metadata matrix of items in domain $A$ as $\mathcal{T}^A \!\in\! \mathbb{R}^{|\mathcal{I}^A|}$. Given user behavior sequences from diverse domains, our CPRec can gain a deeper comprehension of the underlying user preference patterns via knowledge adaptation, enabling it to accurately predict the appropriate items $i^X_{q+1}$ with the sequential behaviors $S^{X}_u\!=\!\{i^X_1, i^X_2, \cdots, i^X_q\}$ of user $u$ in the downstream domain $X$.

\subsection{Overview}
\begin{figure}[!t]
    \centering
    \includegraphics[width=1.0\linewidth]{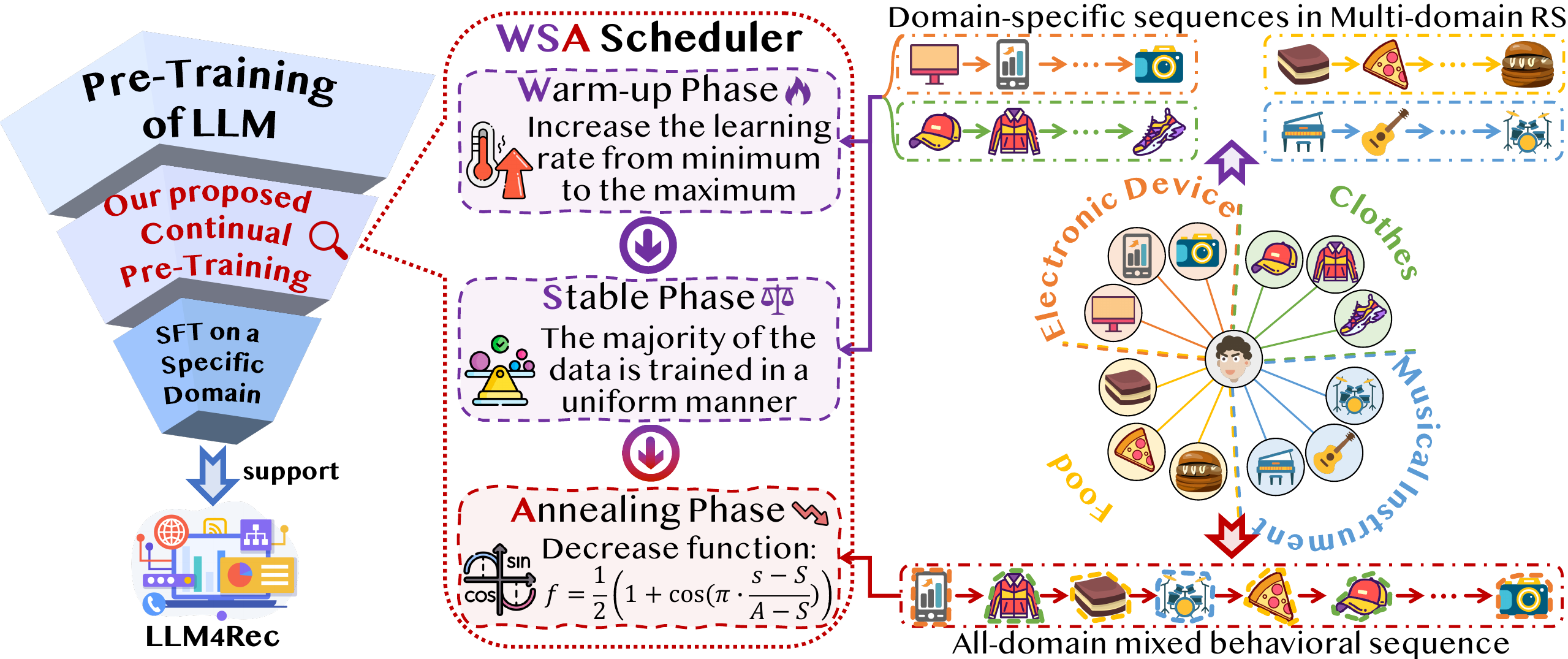}
    \vspace{-0.8cm}
    \caption{The overall structure of CPRec, which serves as the bridge between the pre-training of LLM and the SFT on a specific domain. It incorporates multiple single-domain behavioral data and all-domain mixed data into training via WSA scheduler for seamless modality adaptation.}
    \vspace{-0.5cm}
    \label{fig.overall_structure}
\end{figure}

We elaborate a LLM-enhanced recommendation framework, CPRec, which leverages all-domain continual pre-training to force the semantic modality within pre-trained LLM to imitate the general user behavioral patterns, thereby fully leverage its complex understanding capability to mine the domain-specific collaborative modality on diverse domains. 
As shown in Figure ~\ref{fig.overall_structure}, we first deliberate on the types of behavioral data that are necessary for all-domain continual pre-training. Subsequently, we re-review the prompt format in existing LLM4Rec works, opting for a more generalized template that aligns more closely with the natural language. Finally, we investigate a tailored learning rate scheduler for the all-domain continual pre-training paradigm, ensuring that the LLM retains its robust linguistic capabilities while effectively modeling users' sequential preference patterns. 

\subsection{Behavioral Corpus Structuralization} 
The pre-training of LLMs from scratch necessitates the aggregation of vast data from extensive sources, encompassing both the general text data and the specialized data \cite{LLM_survey}. This implies that LLMs have gained a wealth of open-world knowledge from the semantic modality and possess the potential to handle diverse downstream tasks. To adapt it into the collaborative modality in a continual pre-training manner, we propose a behavioral corpus structuralization strategy, which \emph{consistently adheres to the principle of being ``\textbf{general}''}. It systematically gathers user behaviors from publicly available datasets spanning multiple domains, coverts these massive behaviors into the domain-specific behavioral sequences and all-domain mixed behavioral sequences, and employs a unified prompt format to re-frame users' discrete behaviors in the form of natural language comprehensible to LLMs. 

\vspace{-0.15cm}
\begin{tcolorbox}[top=0.3mm, bottom=0.3mm, left=1mm, right=1mm] 
\textbf{Guidelines of mixed behavior sequence construction}:
\vspace{-2mm} 
\tcblower
\vspace{-2mm} 
1. The domain of the target item to be predicted \textbf{must} have been present in the historical sequence.\\
2. The items of the historical sequence \textbf{must} come from at least two distinct domains.
\end{tcolorbox}
\vspace{-0.15cm}
\subsubsection{Domain-specific Data Structuralization.}

To enable the adaptation from the semantic modality of LLMs into the collaborative modality of the recommendation tasks,
we first construct the domain-specific user behavioral sequences. Specifically, we extensively collect the massive user behaviors from multiple available domains as the continual pre-training corpus, which has been well-verified by several multi-domain recommenders \cite{Recformer, UniSRec}. It can guarantee LLMs capture the mainstream behavioral patterns of the majority, thereby enhancing their generalization in unseen domains. More importantly, integrating user behaviors from multiple domains not only amplifies the training instances in CPT but also prevents LLMs collapse in specific domains, which is analogous to the prevalent pre-training phase of LLMs.

\subsubsection{Mixed-domain Data Structuralization.}
Given the domain-specific behavioral sequences from multiple domains, we innovatively propose the all-domain mixed behavioral sequences. It serves as the complete user behavioral sequence containing all behaviors of the same user from all of the available domains, which represents the authentic user action logic. Explicitly modeling such information in CPT would be beneficial in mining the general user behavioral pattern. Given domain $A$, $B$ and $C$ as an example, with the behavior sequences $S^{A}_u\!=\!\{\!i^A_1, i^A_2, \cdots, i^A_{p^A}\}$, $S^{B}_u\!=\!\{\!i^B_1, i^B_2, \cdots, i^B_{p^B}\}$ and $S^{C}_u\!=\!\{\!i^C_1, i^C_2, \cdots, i^C_{p^C}\}$ from diverse domains of the same user $u$, we concatenate and re-rank them in the chronological order as $S^{M}_u\!=\!\{\!i^B_1, i^C_2, i^A_3, \cdots, i^A_{\!p^A+p^B+p^C\!}\}$, where $p^A$, $p^B$ and $p^C$ denote this user's behavior length in domain $A$, $B$ and $C$ respectively. 
Besides, we also introduce a guideline to outline the construction rules for mixed behavioral sequences in the following text box, thereby obtaining the final mixed behavioral sequences for the following CPT procedure.

These two guidelines jointly ensure that the mixed sequences exhibit strong relevance to the target item, thereby facilitating the extraction of general user behavioral patterns. Specifically, this type of behavioral sequence can represent the complete preference evolutionary pattern and highlight the cross-domain preference correlations simultaneously. Notably, the CPT process of our CPRec is entirely independent of individual users. Thus, we do not necessitate the user overlap constraints imposed by existing cross-domain recommenders like C$^2$DSR \cite{C2DSR} and Tri-CDR \cite{Tri-CDR}, guaranteeing the enhanced generalization ability while remaining unaffected by the sparsity issue within different datasets.

\subsubsection{Unified Prompt Format Design}
Different from the conventional recommenders that leverage the discrete identities to represent users and items, LLM4Rec opt to expand the complex reasoning ability of LLMs to user preference understanding tasks via instruction tuning paradigm. To achieve this, they typically define various prompt templates to guarantee LLMs exhibit remarkable instruction-following ability in recommendation scenarios. Therefore, the prompt templates utilized in these approaches are highly intertwined with the domain-specific information of the specific recommendation scenario, making it challenging to adapt to our all-domain continual pre-training flexibly.

In response to this observation, we introduce a straightforward yet universal prompt template to safeguard the prompt tuning process of LLMs from being influenced by the domain details and the quality of the pre-extracted collaborative signals. Specifically, given the substantial disparity between user behaviors and the syntax in natural language, we believe that \emph{a purely-textual prompt offers a promising approach to bridge the semantic understanding ability of LLMs with the collaborative modality in recommendation objectives} and seamlessly convert the next-token prediction task in LLMs into the next-item prediction task in recommendation. Therefore, we remove the specific collaborative token $\bm{e}^c$ in LLaRA's prompt and then replace the representative name of each domain with the general word ``item'' to enable the CPT phase to be in harmony with the multiple domains. It also guarantees the scalability of our CPRec, allowing the incorporation of behavioral data from diverse domains and platforms as the computational resources expand while alleviating the issue of ``Catastrophic Forgetting'' to some extent.

\subsection{Continual Pre-training Scheduler}

Intuitively, the domain-specific and the all-domain mixed behavioral sequences respectively assist LLMs in grasping user preference evolution patterns within the single domain and aligning with the logical chain of users' real-world preference dynamics, with the latter posing a substantially bigger challenge than the former. 
Therefore, indiscriminately incorporating all data in CPT may result in computational wastage and confused optimization direction. In contrast, strategically introducing varied data types across distinct training phases is a more rational manner to facilitate LLMs converge toward the global minima efficiently.

\begin{table}
\centering
\caption{An example of our unified prompt template, where $t_n$ denotes the textual metadata of the $n$-th item.}
\vspace{-0.3cm}
\label{tab:prompt}
    \begin{tabular}{p{7.7cm}}  
    \toprule
        This user has bought $t_1$, $t_2$, ..., $t_m$ in the previous. Please predict the next item this user will buy. The item title candidates are $t_a$, $t_b$, ..., $t_n$. Choose only one item from the candidates. The answer is \\
    \bottomrule
    \end{tabular}
\vspace{-0.4cm}
\end{table}

Inspired by the systematic analysis of the Cosine Learning Rate Scheduler (LRS) and Warmup-Stable-Decay LRS in MiniCPM \cite{MiniCPM}, we propose the Warmup-Stable-Annealing (WSA) LRS, which can be elegantly integrated with diverse types of behavioral data and diverse hardness of the preference understanding task. As its name implies, WSA LRS can be divided into three explicitly-defined stages: warm-up stage, stable stage, and annealing stage, enabling tailored application of varied data types. Given the end step of the above stages as $W$, $S$ and $A$, the overall learning rate arrangement $\operatorname{WSA}(s)$ at step $s$ can be formulated as:

\begin{eqnarray}
\vspace{-0.1cm}
\operatorname{WSA}(s)= 
    \begin{cases}
        \eta_l + \frac{s}{W} (\eta_h-\eta_l), & s \leq W \\
        \eta_h, & W < s \leq S \\
        \eta_l + f(s)(\eta_h-\eta_l), & S < s \leq A
    \end{cases}
\vspace{-0.1cm}
\end{eqnarray}
where $f(s) \!\in\! [0,\!1]$ is a decreasing function about $s$, $\eta_l$ and $\eta_h$ denote the minimum and the maximum learning rate respectively. 

\subsubsection{Warm-up Phase}
To facilitate the healthy convergence during the initial training phase and preserve the language proficiency of LLMs throughout CPT, we refer to the warm-up strategy \cite{Gradual_warmup} to gradually ramp up the learning rate from a small value to the regular one. Here, the magnitude of the step number $W$ of the warm-up phase can determine its acceleration. Notably, we exclusively leverage the domain-specific user behavioral sequences for this phase. This enables LLMs that are pre-trained in the massive textual information to smoothly adapt from their innate completion ability to user preference understanding tasks.

\subsubsection{Stable Phase}
The stable phase serves as the most extended training phase of CPT with most of the training instances, where we assign the fixed learning rate to guarantee that the majority of behavioral data is uniformly and comprehensively trained. Similar to the warm-up phase, this phase also regards domain-specific behavioral data as its input source. This approach enables the free and sustainable integration of behavioral data from new users or diverse platforms into the CPT process with sufficient computational power and data availability in the future. Therefore, it significantly improves the scalability of CPRec and converts the concept of ``continual'' deployment into an achievable goal. 

\subsubsection{Annealing Phase}
\label{sec.schedule_annealing}
We believe that the pre-trained LLM acquires the capacity to understand the domain-specific user preferences after the warm-up and the stable phase. Regarding the annealing phase, we incorporate the more complex and more authentic all-domain mixed behavioral data to enable the LLMs to capture more generalized user behavioral patterns. To achieve this, we use the cosine scheduler to fit the learning rate annealing process, which has been well-verified in Chinchilla \cite{Chinchilla} and Gopher \cite{Gopher}. The decreasing function $f(s)$ can be formulated as:
\vspace{-0.2cm}
\begin{equation}
    f(s)=\frac{1}{2}(1+\operatorname{cos}(\pi \cdot \frac{s-S}{A-S})),
\end{equation}
where $s$ is the current step. This decreasing method enables the longer duration of the higher learning rate and the smooth decline from maximum to minimum, facilitating the LLMs to find the global optimum. With the all-domain continual pre-trained checkpoint, we utilize it to initialize the low-rank decomposition matrices of LLM and subsequently employ SFT on downstream datasets, following the same training procedure of our base LLM-enhanced recommender. Therefore, \emph{the time complexity of CPRec in the downstream datasets is comparable in magnitude to that of LLaRA} (see Section~\ref{sec.computational_complexity_analysis}). This design ensures that LLM prioritizes modeling specific user preferences derived from its general preference understanding ability rather than the open-world semantic knowledge, which is unrelated to RS. Besides, we also implement several straightforward schedulers to demonstrate the effectiveness and robustness of our WSA LRS (see Section~\ref{sec.ablation_study}).

\subsection{Discussion}
Inspired by the textual reasoning ability of LLMs, recent works have started exploring its potential in recommendation, among which Recformer \cite{Recformer} and LLaRA \cite{LLaRA} are most closely aligned with our CPRec. Here, Recformer~\cite{Recformer} adopts bi-directional pre-trained Longformer to encode the plaintext behavioral sequence. However, it solely focuses on the contrastive alignment of intra-domain behaviors, overlooking the informative mixed behavioral sequences, which precisely reflect users' authentic behavioral logic. Besides, Recformer fails to thoroughly explore the design of pre-training scheduler, and the utilized language model is relatively smaller than the prevalent LLMs. LLaRA~\cite{LLaRA} represents the SOTA LLM-enhanced recommender, which proposes a hybrid prompt method to incorporate collaborative features with items' textual descriptions via a curriculum prompt tuning strategy. Nevertheless, it serves as an SFT approach tailored to the specific domain with the well-designed prompts. This design undermines its robustness and generalization across domains.

In contrast, our CPRec systematically investigates the drawbacks within the existing LLM4Rec studies, verifies the feasibility of our CPT paradigm in LLM-enhance recommendation and explores the issues such as data mixing and data scale in recommendation, which are the critical concerns in CPT. Meanwhile, we introduce the behavioral corpus structuralization method and the tailored CPT scheduler to smoothly adapt the open-world knowledge within the universal semantic modality into the specific collaborative modality via the general user behavioral patterns. 
\begin{table}[!t]
\caption{The detailed statistics of the datasets used for the continual pre-training and downstream validation.}
\vspace{-0.3cm}
\label{tab:dataset}
\resizebox{\linewidth}{!}{ 
    \begin{tabular}{c|c|cccc}
    \toprule
    \textbf{Version} & \textbf{Dataset} & \textbf{Users} & \textbf{Items} & \textbf{Interactions} & \textbf{Sparsity} \\ \midrule
    \multirow{7}{*}{\textbf{\begin{tabular}[c]{@{}c@{}}Continual \\ Pre-training\end{tabular}}}       
        & \textbf{Automotive} & 31,740 & 17,538 & 206,941 & 99.96\% \\
        & \textbf{Clothing} & 35,869 & 18,672 & 234,963 & 99.96\% \\
        & \textbf{Electronic} & 29,248 & 14,450 & 191,680 & 99.95\% \\
        & \textbf{Food} & 35,744 & 15,459 & 243,390 & 99.96\% \\
        & \textbf{Home} & 32,430 & 17,406 & 207,739 & 99.96\% \\
        & \textbf{Movie} & 25,431 & 12,588 & 203,730 & 99.94\% \\ 
        & \textbf{Phone} & 18,500 & 6,497  & 95,483  & 99.92\% \\ \midrule
    \multirow{5}{*}{\textbf{\begin{tabular}[c]{@{}c@{}}Downstream \\ Validation\end{tabular}}} 
        & \textbf{Game}          & 4,564  & 3,058  & 30,302  & 99.78\% \\
        & \textbf{Music}       & 7,501  & 3,547  & 48,240  & 99.82\% \\
        & \textbf{Industry}    & 5,526  & 2,739  & 29,325  & 99.81\% \\
        & \textbf{Bili\_Movie}   & 7,971  & 1,390  & 42,936  & 99.61\% \\ 
        & \textbf{Bili\_Cartoon} & 6,037  & 2,115  & 36,954  & 99.71\% \\ \bottomrule
    \end{tabular}}
\vspace{-0.4cm}
\end{table}

\section{Experiments}\label{sec:experiments}
We conduct extensive experiments to answer the following research questions:\\
\noindent
\textbf{RQ1}: How does CPRec perform against existing sequential recommenders and LLM-enhanced recommenders? (see Section~\ref{sec.performance_comparison}) \\
\noindent
\textbf{RQ2}: How does each component in CPRec impact the downstream recommendation performance? (see Section~\ref{sec.ablation_study})\\
\noindent
\textbf{RQ3}: Can CPRec comprehend general user preferences across scenarios that vary in sparsity? (see Section~\ref{sec.universality_analyses})\\
\noindent
\textbf{RQ4}: How does CPRec function on diverse evaluation metrics, training scales and computational efficiency? (see Section~\ref{sec.in_depth_analyses})

\begin{table*}[!t]
\caption{Performance comparison on downstream datasets. The best and second-best results of each dataset are in bold and underlined, respectively. $^{*}$ denotes significant improvements of CPRec over the baselines (\emph{p} $\textless$ 0.05 with paired t-tests).}
\vspace{-0.3cm}
\label{tab:performance_comparison}
\resizebox{\linewidth}{!}{ 
\begin{tabular}{c|cc|cc|cc|cc|cc}
\toprule
\multirow{2}{*}{\textbf{Algorithms}} &
  \multicolumn{2}{c|}{\textbf{Game}} &
  \multicolumn{2}{c|}{\textbf{Music}} &
  \multicolumn{2}{c|}{\textbf{Industry}} &
  \multicolumn{2}{c|}{\textbf{Bili\_Movie}} &
  \multicolumn{2}{c}{\textbf{Bili\_Cartoon}} \\ 
    & HR@1   & ValidRatio & HR@1    & ValidRatio & HR@1   & ValidRatio & HR@1   & ValidRatio & HR@1   & ValidRatio \\ \midrule
\textbf{GRU4Rec} & 0.3063 & 1.0000 & 0.3371  & 1.0000 & 0.2846 & 1.0000 & 0.2283 & 1.0000 & 0.2401 & 1.0000  \\
\textbf{SASRec}  & 0.3443 & 1.0000 & 0.3530  & 1.0000 & 0.3096 & 1.0000 & 0.2497 & 1.0000 & 0.2631 & 1.0000 \\
\textbf{CL4SRec} & 0.3482 & 1.0000 & 0.3596  & 1.0000 & 0.3048 & 1.0000 & 0.2555 & 1.0000 & 0.2646 & 1.0000 \\ 
\textbf{TedRec} & 0.3576 & 1.0000 & 0.3798 & 1.0000 & 0.3230 & 1.0000 & \underline{0.2561} & 1.0000 & 0.2771 & 1.0000 \\ \midrule
\textbf{LLaMA-2} & 0.0741 & 0.4912 & 0.0252 & 0.2570 & 0.0460 & 0.3230 & 0.0238 & 0.4628 & 0.0209 & 0.4509 \\
\textbf{ChatRec} & 0.2382 & 0.9327 & 0.1274  & 0.9367 & 0.2302 & 0.8189 & 0.0544 & 0.8574 & 0.1135 & 0.9289 \\
\textbf{MoRec}   & 0.3857 & 1.0000 & 0.3773  & 1.0000 & 0.3460 & 1.0000 & 0.2285 & 1.0000 & 0.2575 & 1.0000 \\
\textbf{LLaRA}   & \underline{0.4411} & 0.9851 & \underline{0.3972} & 0.9971 & \underline{0.3577} & 0.9860 & 0.2532 & 0.9982 & \underline{0.2816} & 0.9988 \\ \midrule
\textbf{CPT-LLaRA}& 0.3775 & 0.9779 & 0.2562 & 0.8910 & 0.2856 & 0.8823 & 0.0629 & 0.8883 & 0.1280 & 0.8814 \\
\cellcolor{gray!16}\textbf{CPRec} & \cellcolor{gray!16}\textbf{0.4764$^{*}$} & \cellcolor{gray!16}0.9952 & \cellcolor{gray!16}\textbf{0.4286$^{*}$} & \cellcolor{gray!16}0.9912 & \cellcolor{gray!16}\textbf{0.3730$^{*}$}  & \cellcolor{gray!16}0.9788& \cellcolor{gray!16}\textbf{0.2706$^{*}$} & \cellcolor{gray!16}0.9950 & \cellcolor{gray!16}\textbf{0.3129$^{*}$} & \cellcolor{gray!16}0.9906 \\ \midrule
\emph{\textbf{Improvement}} & \color{red}+8.00\% & - & \color{red}+7.91\% & - & \color{red}+4.28\% & - & \color{red}+5.66\% & - & \color{red}+11.12\% & - \\ \bottomrule
\end{tabular}}
\vspace{-0.4cm}
\end{table*}

\subsection{Experiment Setup}

\subsubsection{Datasets.}
This paper focuses on the all-domain continual pre-training for recommendation, which is inherently structured into two distinct stages: continual pre-training and downstream validation.
Specifically, we collect behavioral data from seven domains in Amazon \cite{Amazon} for the former and construct five datasets that are unseen in CPT from two platforms, Amazon \cite{Amazon} and Bilibili \cite{Bilibili}, for the latter to demonstrate our effectiveness and universality. 
Following \cite{PDRec,SeeDRec,RealHNS}, we adopt 3-core filter and Leave-one-out method to construct each user's behavioral sequence in the chronological order, regarding the last behavior for testing, the penultimate one for validation, and other behaviors for training. The detailed statistics are shown in Table \ref{tab:dataset}~\footnote{Due to the limited page, we detail other statistics of our datasets and their construction protocols in Appendix.}.

\subsubsection{Baseline Methods.}
To assess the effectiveness of CPRec, we compare it with four traditional recommenders (\eg GRU4Rec \cite{GRU4Rec}, SASRec \cite{SASRec}, CL4SRec \cite{CL4SRec} and TedRec \cite{TedRec}) and five LLM-enhanced recommenders (\eg LLaMA-2 \cite{Llama2}, ChatRec \cite{ChatRec}, MoRec \cite{MoRec}, LLaRA \cite{LLaRA} and  CPT-LLaRA). More details are in Appendix.

\subsubsection{Implementation Details.}
We implement all the methods with Python 3.10.14 on one NVIDIA H100 and take Adam as the optimizer. For the entire CPT process, we employ LoRA \cite{LoRA} which leverages the low-rank decomposition matrices to reduce the number of necessary parameters for modality adaptation. Similar to LLaRA~\cite{LLaRA}, we leverage LLaMA2-7B \cite{Llama2} as the LLM backbone, conduct CPT for only one epoch, configure the number of warm-up samples $W$ as the 5\% of the whole domain-specific user behaviors and define the minimum and maximum learning rate as $5e^{-6}$ and $5e^{-4}$ respectively. Besides, we search the optimal learning rate from $\{6e^{-4}, 3e^{-4}, 1e^{-4}\}$ for downstream validation. To alleviate randomness, we conduct four runs with different seeds and report the average results for all models~\footnote{The implementation details of traditional recommenders and LLM-enhanced recommenders are listed in Appendix.}.

\subsubsection{Evaluation Metrics.}
Given that CPRec pertains to the all-domain continual pre-training in RS, we evaluate its performance in several downstream recommendation datasets. Following \cite{LLaRA, S-DPO, RosePO}, we randomly sample 19 non-interacted items for each positive instance and select the Hit Ratio@1 (HR@1) and the Valid Ratio to evaluate its performance and its instruction following ability in generating the appropriate items. Besides, we additionally conduct the universality analysis of our CPRec and some representative baselines on HR@k and Normalized Discounted Cumulative Gain (N@k) with different $k \in \{1, 3, 5\}$ \cite{NS_Survey,DieT}, the further details have been discussed in Section~\ref{sec.evaluation_metrics}.

\subsection{Performance Comparison (RQ1)}
\label{sec.performance_comparison}
We compare CPRec with several recommendation algorithms to indicate its effectiveness. From Table~\ref{tab:performance_comparison}, we can observe that:

(1) CPRec surpasses all the compared baselines on all datasets with the significance level \emph{p} $\!\textless\!$ 0.05. This shows that CPRec can capture the intermediate general user behavioral patterns, enabling the smooth adaptation from the open-world semantic knowledge inherent in LLMs into domain-specific collaborative modality.

(2) Comparing the performance among different datasets, some LLM-enhanced recommenders perform worse than the traditional recommenders in Bili-series datasets, which may due to the uneven semantic qualities and the interest distributions across distinct platforms. In contrast, our CPRec exhibits more stable and more superior performance than these two types of algorithms across different downstream domains and repeated trials, indicating its stronger robustness. More important, these Bili-series datasets differ from the data source (Amazon) employed in our CPT process. This highlights the universality of our CPRec, demonstrating its adaptability to unseen datasets with varying user behavioral distributions from diverse platforms. 

(3) CPT-LLaRA achieves significant performance improvement over the traditional recommenders on Game. This validates CPT's ability to extract the general preference patterns from users' domain-specific and all-domain mixed behavioral sequences, thereby offering valuable information gains to support existing LLM4Rec works (more details can be found in Section~\ref{sec.zero_shot_model_analyses}).

\begin{figure*}[!t]
    \centering
    \includegraphics[width=0.98\linewidth]{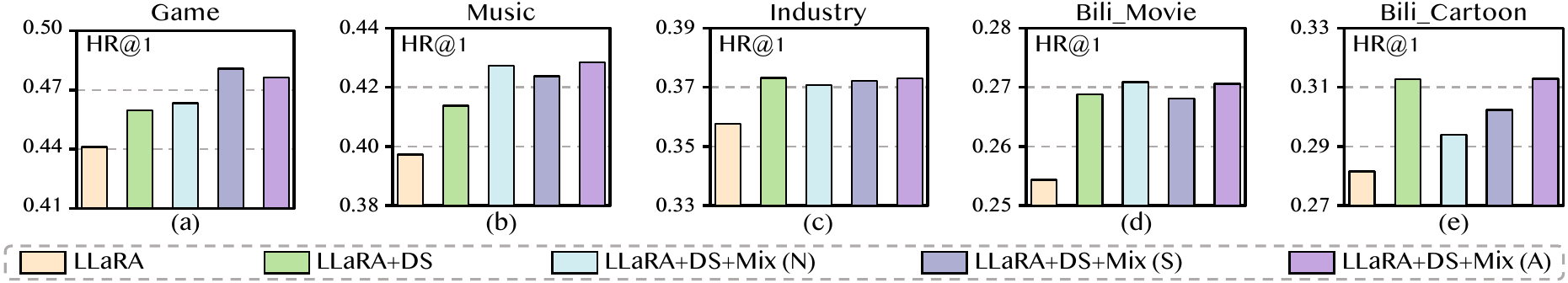}
    \vspace{-0.3cm}
    \caption{Results of CPRec and its ablation versions on datasets from two diverse platforms. All components are effective.}
    \vspace{-0.4cm}
    \label{fig.ablation_study}
\end{figure*}

\subsection{Ablation Study (RQ2)}
\label{sec.ablation_study}
To investigate the effectiveness of each component in CPRec, we compare it with LLaRA, LLaRA+DS, LLaRA+DS+Mix (N) and LLaRA+DS+Mix (S). Here, LLaRA+DS denotes solely employing CPT with the domain-specific behavioral data, LLaRA+DS+Mix (N) denotes indiscriminately modeling the domain-specific behavioral data and the all-domain mixed behavioral data in CPT, LLaRA+DS+Mix (S) denotes replacing the annealing phase in WSA scheduler with the stable phase, and LLaRA+DS+Mix (A) equals CPRec. From Figure~\ref{fig.ablation_study}, we have the following observations:

(1) LLaRA+DS consistently outperforms LLaRA on all datasets. That is, even a simple implementation achieves significant performance gains, implying the potential of CPT in recommendation. 

(2) LLaRA+DS+Mix (N) exhibits improved performance over LLaRA+DS on Game, Music and Bili\_Movie, validating the effectiveness of all-domain mixed behavioral data in CPT. It is primarily due to the fact that it reflects real-world user behavioral logic, and including it in CPT can significantly strengthen LLMs' capacity in completely understanding general user preferences.

(3) Comparing LLaRA+DS+Mix (A) with LLaRA+DS+Mix (N), we reconfirm that the all-domain mixed behavioral sequences are informative, while directly incorporating it into the CPT process might introduce extra noise. Tailored optimization strategies should be investigated to enable fine-grained modeling.

(4) In general, LLaRA+DS+Mix (A) attains the peak results across most of the five datasets, which demonstrates its effectiveness and robustness. To precisely align with the pre-training procedure of LLMs and seamlessly integrate each type of user behaviors into CPT, we opt for this implementation as our CPRec.

\begin{table}[!t]
\caption{Performance comparison of the zero-shot analysis from the perspective of recommenders.}
\vspace{-0.3cm}
\label{tab:sft_analysis}
\resizebox{\linewidth}{!}{ 
\begin{tabular}{c|cccc}
\toprule
\textbf{Datasets} & \textbf{SASRec} & \textbf{LLaMA-2} & \textbf{\begin{tabular}[c]{@{}c@{}}LLaRA\\w/o SFT\end{tabular}} & \textbf{\begin{tabular}[c]{@{}c@{}}CPRec\\w/o SFT\end{tabular}} \\ \midrule
Game          & \underline{0.3443} & 0.0741 & 0.0210 & \textbf{0.3775} \\ 
Music         & \textbf{0.3530} & 0.0252 & 0.0120 & \underline{0.2562}          \\ 
Industry      & \textbf{0.3096} & 0.0460 & 0.0158 & \underline{0.2856}          \\ 
Bili\_Movie   & \textbf{0.2497} & 0.0238 & 0.0040 & \underline{0.0629}          \\ 
Bili\_Cartoon & \textbf{0.2631} & 0.0209 & 0.0043 & \underline{0.1280}          \\ \bottomrule
\end{tabular}}
\vspace{-0.4cm}
\end{table}

\subsection{Universality Analysis (RQ3)}
\label{sec.universality_analyses}
\subsubsection{Universality analysis of the zero-shot performance from the perspective of recommenders.}
\label{sec.zero_shot_model_analyses}

To inspect the zero-shot performance of CPRec, we compare CPRec w/o SFT with SASRec, LLaMA-2, and LLaRA w/o SFT. Here, SASRec has undergone the comprehensive training process, whereas the remaining three approaches solely leverage the pre-trained LLMs to infer user preferences without any SFT process. Notably, CPRec w/o SFT involves initializing the low-rank decomposition matrices of LLaRA w/o SFT with the checkpoint derived from our CPT process.

\begin{table}[!t]
\caption{Results of universality analysis on different number of user behaviors.}
\vspace{-0.3cm}
\label{tab:diverse_density}
\resizebox{\linewidth}{!}{ 
\begin{tabular}{c|c|cccc}
\toprule
\textbf{Datasets} & \textbf{Methods} & \textbf{Sparse} & \textbf{Medium} & \textbf{Dense} & \textbf{All} \\ \midrule
\multirow{2}{*}{Game}   
& LLaRA      & 0.4500        & 0.4264           & 0.4315    & 0.4411       \\
& CPRec      & \textbf{0.4849}           & \textbf{0.4661}           & \textbf{0.4588} & \textbf{0.4764}\\ \midrule
\multirow{2}{*}{Music}    
& LLaRA      & 0.4370           & 0.3612           & 0.2921    &  0.3972       \\
& CPRec      & \textbf{0.4648}           & \textbf{0.3945}           & \textbf{0.3354}   &  \textbf{0.4286} \\ \midrule
\multirow{2}{*}{Industry} 
& LLaRA      & 0.3711           & 0.3193           & 0.3370    &    0.3577     \\
& CPRec      & \textbf{0.3853}           & \textbf{0.3361}           & \textbf{0.3608}  & \textbf{0.3730}\\ \midrule
\multirow{2}{*}{Bili\_Movie} 
& LLaRA      & 0.2516           & 0.2576           & 0.2530      &   0.2532    \\
& CPRec      & \textbf{0.2685}           & \textbf{0.2772}           & \textbf{0.2657}  & \textbf{0.2706} \\ \midrule
\multirow{2}{*}{Bili\_Cartoon} 
& LLaRA      & 0.2704           & 0.3034           & 0.2724      &   0.2816    \\
& CPRec      & \textbf{0.3022}           & \textbf{0.3262}           & \textbf{0.3276}  & \textbf{0.3129} \\ \bottomrule
\end{tabular}}
\vspace{-0.5cm}
\end{table}

\begin{table*}[!t]
\caption{Results of robustness analysis on the diverse evaluation metrics with $k \in \{1, 3, 5\}$.}
\vspace{-0.3cm}
\label{tab:robustness_analyses}
\resizebox{\linewidth}{!}{ 
\begin{tabular}{c|ccccc|ccccc|ccccc}
\toprule
\textbf{Datasets} & \multicolumn{5}{c|}{\textbf{Game}} & \multicolumn{5}{c|}{\textbf{Music}} & \multicolumn{5}{c}{\textbf{Industry}} \\ \midrule
\textbf{Metric} & HR@1 & HR@3 & HR@5 & N@3 & N@5 & HR@1 & HR@3 & HR@5 & N@3 & N@5 & HR@1 & HR@3 & HR@5 & N@3 & N@5 \\ \midrule
\textbf{SASRec} & 0.3443 & 0.5552 & 0.6612 & 0.4667 & 0.5104 & 0.3530 & 0.5298 & 0.6302 & 0.4553 & 0.4966 & 0.3096 & 0.4717 & 0.5612 & 0.4033 & 0.4401 \\
\textbf{LLaMA-2} & 0.0758 & 0.1573 & 0.2314 & 0.1227 & 0.1529 & 0.0655 & 0.0265 & 0.1007 & 0.0489 & 0.0633 & 0.1082 & 0.0530 & 0.1587 & 0.0847 & 0.1052 \\
\textbf{ChatRec} & 0.2382 & 0.4266 & 0.5173 & 0.3472 & 0.3846 & 0.1274 & 0.2896 & 0.4135 & 0.2198 & 0.2706 & 0.2302 & 0.3668 & 0.4394 & 0.3092 & 0.3392 \\
\textbf{LLaRA}  & \underline{0.4411} & \underline{0.6460} & \underline{0.7333} & \underline{0.5614} & \underline{0.5973} & \underline{0.3972} & \underline{0.5832} & \textbf{0.6753} & \underline{0.5055} & \underline{0.5434} & \underline{0.3577} & \underline{0.5315} & \underline{0.6188} & \underline{0.4589} & \underline{0.4947} \\
\textbf{CPRec}  & \textbf{0.4764} & \textbf{0.6761} & \textbf{0.7422} & \textbf{0.5937} & \textbf{0.6210} & \textbf{0.4286} & \textbf{0.5843} & \underline{0.6505} & \textbf{0.5196} & \textbf{0.5469} & \textbf{0.3730} & \textbf{0.5485} & \textbf{0.6350} & \textbf{0.4753} & \textbf{0.5108} \\ \bottomrule
\end{tabular}}
\vspace{-0.2cm}
\end{table*}

\begin{figure*}[!t]
    \centering
    \includegraphics[width=0.98\linewidth]{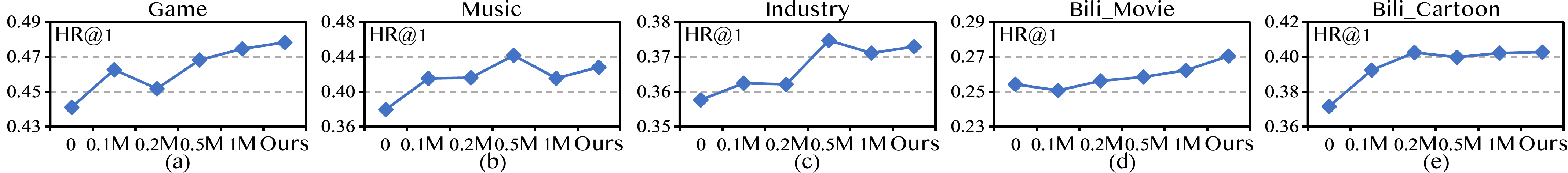}
    \vspace{-0.3cm}
    \caption{Impact of diverse training data volume on downstream datasets. Generally, more pre-training data is effective.}
    \vspace{-0.4cm}
    \label{fig.scaling_analysis}
\end{figure*}

From Table~\ref{tab:sft_analysis}, we notice that:
(1) CPRec w/o SFT consistently exhibits the best and the second-best performance on each dataset without any time-consuming SFT process, which indicates that CPT can effectively encode the general user behavioral patterns from massive user behaviors.
(2) Comparing the performance between SASRec and CPRec w/o SFT on two platforms, CPRec w/o SFT exhibits higher performance on datasets from Amazon compared to that of Bilibili. This discovery is natural as the training instances of CPT exclusively originate from Amazon, which inherently demonstrates a stronger alignment with the behavioral logic of Amazon users. Moreover, it implicitly showcases that our CPT can capture the general preference patterns for diverse sources of users to some extent.
(3) Considering the performance of LLaMA-2 and CPRec w/o SFT across diverse datasets, CPRec w/o SFT demonstrates superior performance to SASRec on datasets where LLaMA-2 excels. This indicates that our CPRec can effectively leverage the semantic reasoning ability of the LLM backbone and smoothly adapt it into the collaborative modality understanding task. Furthermore, it confirms that CPRec can fully leverage the possible advancements of LLMs in the future, thereby prolonging the lifecycle of our proposed all-domain continual pre-training paradigm.

\subsubsection{Universality analysis among diverse sparsity of behaviors}
To evaluate the universality of CPRec on diverse sparsity of behaviors, we divide users into three subsets: \emph{sparse}, \emph{medium}, and \emph{dense}, during the inference phase and report the performance of LLaRA and CPRec on them respectively. As for the specific setting in LLaRA \cite{LLaRA} and the Leave-one-out strategy, the number of user historical behaviors spans from 3 to 10 for the inference phase. Thus, we designate 5 and 9 as the thresholds to divide the above three subsets with the left-closed and right-open interval. 

Table~\ref{tab:diverse_density} shows that CPRec significantly surpasses LLaRA on all subsets and all datasets, which highlights the universality of our CPRec in different recommendation scenarios with diverse sparsity of user behaviors. 
As for the performance on different subsets of the same dataset, CPRec achieves optimal results on the \emph{sparse} subsets of Amazon datasets and the comparable performance on \emph{medium} and \emph{dense} subsets of Bilibili ones. This phenomenon is reasonable as our CPRec is pre-trained on the Amazon datasets, making it adept at capturing users' behavioral patterns specific to that platform with fewer behaviors. In contrast, for the unfamiliar Bilibili datasets, more behaviors ought to be necessary to generalize the preference knowledge effectively.

\subsection{In-depth Model Analysis (RQ4)}
\label{sec.in_depth_analyses}

\subsubsection{Robustness analysis on diverse evaluation metrics}
\label{sec.evaluation_metrics}
To assess the robustness of CPRec on diverse evaluation metrics, we evaluate it with two typical metrics (HR@k and N@k with $k \in \{1, 3, 5\}$). Following \cite{LLaRA}, CPRec aims to generate the most appropriate item with the given candidates, which differs from the objective of top-k recommendation. Thus, we conduct simulation experiments to enable LLM to predict an ordered item sequence. Specifically, once CPRec generates an item, it will be removed from the candidates, and then the generation process will continue. Through five iterations of the above procedure, we can derive the ordered sequence of five items. As illustrated in Table~\ref{tab:robustness_analyses}, we observe that: (1) CPRec exhibits significant performance improvement over all baselines on most datasets. The relative improvements are larger with the smaller k and rank-sensitive metric (N@k). It is reasonable that CPRec can uncover users' genuine behavior logic by explicit modeling of the domain-specific and all-domain mixed behavioral sequences to facilitate the target item to be predicted in the top-ranked positions. (2) The performance of LLaMA-2 and ChatRec which are non-equipped with SFT is generally lower than that of traditional recommenders while CPRec consistently outperforms. This re-highlights the necessity of SFT in LLM-enhanced recommenders and the robustness of our CPRec. 

\begin{table}[!t]
\caption{Computational complexity comparison between LLaRA and CPRec. ``\#Tra.'' and ``\#Inf.'' denote training and inference period, while ``(m)'' and ``(MB)'' denote the running time and the occupied GPU memory for each epoch.}
\vspace{-0.3cm}
\label{tab:computional_cost}
\resizebox{\linewidth}{!}{ 
\begin{tabular}{c|c|cc|cc}
\toprule
\textbf{Datasets} & \textbf{Methods} & \textbf{\#Tra. (m) $\downarrow$} & \textbf{\#Tra. (MB) $\downarrow$} & \textbf{\#Inf. (m) $\downarrow$} & \textbf{\#Inf. (MB) $\downarrow$} \\ \midrule

\multirow{2}{*}{\fontsize{10pt}{11pt}\selectfont Game} & \fontsize{10pt}{11pt}\selectfont LLaRA & \fontsize{10pt}{11pt}\selectfont 27:59 & \fontsize{10pt}{11pt}\selectfont 68,428 & \fontsize{10pt}{11pt}\selectfont 15:19 & \fontsize{10pt}{11pt}\selectfont 45,902 \\
 & \fontsize{10pt}{11pt}\selectfont CPRec & \fontsize{10pt}{11pt}\selectfont 27:55 & \fontsize{10pt}{11pt}\selectfont 68,428 & \fontsize{10pt}{11pt}\selectfont 14:48 & \fontsize{10pt}{11pt}\selectfont 45,902 \\ \midrule
\multirow{2}{*}{\fontsize{10pt}{11pt}\selectfont Music} & \fontsize{10pt}{11pt}\selectfont LLaRA & \fontsize{10pt}{11pt}\selectfont 66:26 & \fontsize{10pt}{11pt}\selectfont 75,074 & \fontsize{10pt}{11pt}\selectfont 29:39 & \fontsize{10pt}{11pt}\selectfont 62,786 \\
 & \fontsize{10pt}{11pt}\selectfont CPRec & \fontsize{10pt}{11pt}\selectfont 66:17 & \fontsize{10pt}{11pt}\selectfont 75,074 & \fontsize{10pt}{11pt}\selectfont 29:26 & \fontsize{10pt}{11pt}\selectfont 62,786 \\ \midrule
\multirow{2}{*}{\fontsize{10pt}{11pt}\selectfont Industry} & \fontsize{10pt}{11pt}\selectfont LLaRA & \fontsize{10pt}{11pt}\selectfont 43:54 & \fontsize{10pt}{11pt}\selectfont 78,858 & \fontsize{10pt}{11pt}\selectfont 26:05 & \fontsize{10pt}{11pt}\selectfont 57,988 \\
 & \fontsize{10pt}{11pt}\selectfont CPRec & \fontsize{10pt}{11pt}\selectfont 43:53 & \fontsize{10pt}{11pt}\selectfont 78,858 & \fontsize{10pt}{11pt}\selectfont 26:17 & \fontsize{10pt}{11pt}\selectfont 57,988 \\ \bottomrule
\end{tabular}}
\vspace{-0.5cm}
\end{table}

\subsubsection{Impact of training data volume on downstream recommendation performance}
\label{sec.data_volume_analyses}
Regarding the CPT paradigm, the most conspicuous issue is how training data volume affects the downstream recommendation performance. To investigate this problem, we leverage several intermediate checkpoints on diverse magnitude of training instances (\eg 0, 0.1M, 0.2M, 0.5M, and 1M, where M is million) as the initialized low-rank decomposition matrices of pre-trained LLMs and subsequently conduct SFT on each dataset. Notably, ``0'' (the training scale of zero) corresponds to LLaRA itself, and ``Ours'' represents the ultimate version of CPRec.

From Figure~\ref{fig.scaling_analysis}, we notice that: (1) The performance curves across five datasets indicate that the performance of CPRec would generally improve with the increase of the training scale in CPT, which aligns with the prior findings in \cite{Scaling_Law_2} and \cite{Scaling_Law_1}. Thus, we have reason to believe that \emph{the current performance just represents its temporary optimum, and CPRec can achieve better results with increased training data volume.} (2) Compared with the ``Ours'' version, CPRec exhibits acceptable performance on certain datasets with relatively small training data volume (\eg 0.5M). This implicitly reflects that \emph{the effectiveness of CPT paradigm is not purely reliant on the accumulation of the training data volume}.

\subsubsection{Computational complexity analysis} 
\label{sec.computational_complexity_analysis}
To verify the lightweight characteristic of CPRec in downstream validation, we conduct the computational complexity analysis on three datasets with the same batch size of LLaRA as 4, 4 and 2. Recent advances in inference acceleration have significantly lowered the deployment barrier of LLMs, enabling their flexible utilization in daily applications~\cite{GPT-4,DeepSeek-R1}. The queries in these applications align closely with those in LLM4Rec, allowing such practical infrastructures to be seamlessly adapted for LLM4Rec with promising performance. More importantly, \emph{CPRec follows a once-training-for-all paradigm and requires only 55 GPU hours on a single H100 for the CPT stage}—an extremely lightweight cost compared to typical LLM applications. Besides, we also discover that CPRec exhibit the asymptotic similar computational complexity with existing LLM4Rec studies in downstream RS scenarios from Table~\ref{tab:computional_cost}. These findings collectively demonstrate the lightweight nature of CPRec and its promising effectiveness in LLM4Rec tasks.
\section{Conclusion}\label{sec:conclusion}
In this paper, we introduced a general recommendation framework, All-domain Continual Pre-Training for Recommendation (CPRec), which innovatively integrates continual pre-training paradigm into existing LLM-enhanced recommenders to enable the smooth modality adaptation from the universal semantic modality to the domain-specific collaborative modality via the general user behavioral patterns. 
With the behavioral corpus structuralization method and the tailored optimization scheduler, CPRec could gradually familiarize itself with the general preference comprehension task in a step-by-step manner. We collected million-level behaviors from several domains for CPT and conducted extensive downstream experiments to verify the effectiveness and robustness of our CPRec on five unseen datasets. 

We believe that our proposed all-domain continual pre-training paradigm will provide a solid foundation for academic researchers and industry developers to explore LLM4Rec. In the future, we intend to incorporate more behavioral data from other platforms into CPT to improve its generalization and robustness. Moreover, we will investigate the prior-training of pre-trained LLMs with the behavior-agnostic data within the recommendation scenarios and incorporate the explicit preference learning via DPO into our CPRec as part of the post-training stage to further strengthen its vertical recommendation capabilities.

\balance
\bibliographystyle{ACM-Reference-Format}
\bibliography{Ref}

\end{sloppypar}
\end{document}